\documentclass[preprint,showpacs,preprintnumbers,amsmath,amssymb]{revtex4}

\usepackage{graphicx}
\usepackage{dcolumn}
\usepackage{bm}

\newcommand{\Buzek}{Bu\v{z}ek}

\newcommand{\half}{\frac{1}{2} }
\newcommand{\real}{\mathbf{R}}

\newcommand{\Hilbert}{\mathcal{H} }

\newcommand{\state}{\mathcal{S}(\mathcal{H})}

\newcommand{\Tr}{\mathrm{Tr}}

\newcommand{\Exp}{\mathrm{E}}

\newcommand{\rmd}{\mathrm{d}}

\newcommand{\dtheta}{\mathrm{d}\theta}
\newcommand{\dx}{\mathrm{d}x}

\newcommand{\proof }{\textit{Proof.}\\}

\newcommand{\qed}{\textit{Q.E.D.}\\}

\begin{document}

\preprint{APS/123-QED}

\title{Generalized Bayesian predictive density operators}

\author{Fuyuhiko Tanaka}
\email{ftanaka@stat.t.u-tokyo.ac.jp}
 \affiliation{%
Department of Mathematical Informatics, University of Tokyo, 7-3-1 Hongo, Bunkyo-ku, Tokyo, 113-8656  Japan
}%

\date{\today}

\begin{abstract}
Recently the quantum Bayesian prediction problem was formulated by Tanaka and Komaki (2005). 
It is shown that Bayesian predictive density operators are the best predictive density operators when we evaluate them 
by using the averaged quantum relative entropy based on a prior distribution.
In the present paper, we adopt the quantum $\alpha$-divergence as a wider class of loss function.
The generalized Bayesian predictive density operator is defined and shown to be  best among all the estimates of the unknown density operator.
\end{abstract}

\pacs{03.67.-a,03.65.Yz}
\maketitle

\section{\label{sec:level1}Introduction}


In classical statistics, the problem of predicting an unobserved variable $y$ by using an observed variable $x$
has been investigated.
Suppose that a parametric model \[
 \mathcal{P} = \{ p(y | \theta ) : \theta \in \Theta \},
\]
 which is a set of probability densities, is given,    
where $\Theta $ is a parameter space.
Random variables $x$ and $y$ are distributed according to the same true probability density
 $p(\cdot | \theta )$ in $\mathcal{P}$.
 We predict the unobserved variable $y$ with a predictive density $\hat{p}(y;x) $ constructed by using the observed 
 variable $x$.
The closeness of the true density $p(y| \theta )$ and a predicted density $\hat{p}(y; x)$ 
is evaluated by using the Kullback-Leibler divergence
\[
 D(p||\hat{p} ) := \int p(y| \theta ) \log \frac{p(y| \theta )}{\hat{p}(y ; x)} \rmd y.
\]
Aitchison~\cite{Ait1975} showed that 
a Bayesian predictive density 
\begin{equation}
p_{\pi}(y | x) := \int_{\Theta} p(y|\theta ) \pi(\theta | x)  \dtheta , \label{eq:KL}
\end{equation}
where $\pi(\theta | x)$ is a posterior distribution,  
is the best predictive density when we evaluate a predictive density $\hat{p}(y; x)$ by  using
 the average Kullback-Leibler divergence $\int \pi (\theta ) \int D(p || \hat{p}) p(x | \theta ) \dx \dtheta $, where $\pi (\theta )$ is a probability density.
This result was extended to the quantum setting by Tanaka and Komaki~\cite{FT3}.

Let us consider this result more deeply.
From an observation $x$, we obtain the corrected information $\pi(\theta |x) $ on the unknown parameter $\theta$
 and Eq.(\ref{eq:KL}) is obtained by taking mixture of possible probability densities $p(y| \theta)$ with respect to $\pi(\theta | x)$.
 However, there are many ways of taking mixture. 
For example, let $p_{1}(x)$ and $p_{2}(x)$ denote two possible Gaussian distributions.
Then, $p(x) = \half p_{1}(x) + \half p_{2}(x)$ is one possibility.
Another possibility of mixture is given by $\log p'(x) = \half \log p_{1}(x) + \half \log p_{2}(x)$.
The advantage of the latter is that the mixture itself is again a Gaussian distribution.
In classical statistics, such a mixture is often useful and a general class of mixture,  an $\alpha $-mixture~\cite{Amari2005}, can be defined by
\begin{equation}
\begin{matrix}
 p_{\pi}^{(\alpha)}(y | x) &:=& \left\{ \int \{ p(y|\theta) \}^{ \frac{1-\alpha}{2} } \pi(\theta | x) \rmd \theta   \right\}^{ \frac{2}{1- \alpha} }, \\
 p_{\pi}^{(+1)}(y | x)  &:=&  \exp \left\{ \int  \log (p(y|\theta))   \pi(\theta | x) \rmd \theta  \right\}. 
\end{matrix}
 \label{eq:alphamix}
\end{equation}
Corcuera~and~Giummol\`{e}~\cite{Cor1999} showed that the above predictive distribution (\ref{eq:alphamix}), which they called a generalized Bayesian predictive density, is optimal
 under the following $\alpha $-divergence:
\[
D^{(\alpha)}(p||q) := \frac{4}{1 - \alpha^2} \left\{ 1 - \int p(x)^{\frac{1- \alpha}{2} } q(x)^{\frac{1+\alpha}{2} }  \rmd x \right\},
\]
where $\alpha \neq \pm 1$. 
When $\alpha = \pm 1$, it is defined by
\begin{equation}
\begin{array}{ll}
 D^{(-1)}(p||q) &:= \int p(x) \log \left(p(x)/q(x) \right) \rmd x, \\
D^{(+1)}(p||q) &:= D^{(-1)}(q||p).
\end{array}
 \label{eq:alphadiv}
\end{equation}
The $\alpha$-divergence is closely related to the $\alpha$-entropy of R\'{e}nyi~\cite{Renyi1961} and the Chernoff distance~\cite{Cher1952} in the classical information theory.
Our purpose in the present paper is to extend the result obtained by Corcuera~and~Giummol\`{e}~ to the quatum setting.


 In quantum statistics, which was initiated by  Helstrom, Holevo, and other researchers~\cite{Helstrom, Holevo} a quarter century ago,
  the optimal estimation of the parameter of the unknown quantum state has been one of the hot topics over the past several years 
with recent developments of experimental techniques~\cite{Hayashi2005,Barn2003,Paris2004}.
They usually consider the ideal situation with all measurements (described by POVM) allowed and often deal with large sample cases.
 The theoretical limitation on the accuracy of the parameter estimation has been clarified to some extent.
However, in practical situation, we often need to know the density operator describing the unknown quantum state 
rather than the unknown parameter with a given measurement device.
\Buzek \ \textit{et al.}~\cite{Buzek1998} recommended to use Bayesian technique especially when the sample size of experimental data is small.
They proposed to use a posterior state corresponding to a posterior distribution in classical counterparts. 
Tanaka and Komaki~\cite{FT3} formulated the estimation of the unknown density operator  as the quantum prediction problem 
and showed that the posterior state is best among all the estimates of the unknown density operator.


However, the optimality argument depends on the choice of an evaluation function, which is called {\it a loss function} in mathematical statistics.
Here, we adopt a general class of loss fucntion as a quantum counterpart of the classical $\alpha$-divergence (\ref{eq:alphadiv}).
Then, we define the generalized Bayesian predictive density operator and show that it is the optimal density estimate.
Our result includes the previous result obtained by Tanaka and Komaki~\cite{FT3} as a special case.



In the next section, we briefly review our setting, essentially, the same one as in Tanaka and Komaki~\cite{FT3},
except for the choice of a loss function.
In Section 3, we prove our main result. Concluding remarks are described in Section 4.


\section{Preliminary}

We briefly summarize some notations of quantum measurement. 
Let $\Hilbert $ be a separable (possibly infinite dimensional) Hilbert space of a quantum system.
An Hermitian operator $\rho $ on $\Hilbert $ is called a \textit{state} or \textit{density operator} if it satisfies,
\[
\Tr \rho = 1, \quad \rho \geq 0.
\]
We denote the set of all states on $\Hilbert $ as $\state $. 

Let $\Omega $ be a space of all possible outcomes of an experiment (e.g., $\Omega = \real^{n}$)
 and suppose that a $\sigma$-algebra $\mathcal{B}  := \mathcal{B}(\Omega) $ of subsets of $\Omega $ is given. 
An affine map $\mu $ from $\state $ into a set of probability distributions on $\Omega$, $\mathcal{P}$=
 $\{ \mu (\dx )  \}$ is called a \textit{measurement}.
There is a one-to-one correspondence between a measurement
 and a resolution of the identity~\cite{Holevo}.
A map from $\mathcal{B}$ into the set of positive Hermitian operators 
\[ 
 E: B \mapsto E(B),
\]
where $E $ satisfies
\begin{gather}
 E(\phi) = O, E(\Omega )=I, \label{eq:cond1} \\
 E(\cup_{i} B_i ) = \sum_i E(B_i),  \quad B_i \cap B_j = \phi,  \quad  \forall B_i \in \mathcal{B }, \label{eq:cond2}
\end{gather}
is called a \textit{positive operator valued measure} (\textit{POVM}).
Any physical measurement can be represented by a POVM.

Now we describe our setting of state estimation.
Assume that a state $\rho_{\theta} $ on $\Hilbert $ is characterized by an unknown finite-dimensional parameter $\theta \in \Theta \subset  \real^n$.
If $\dim \Hilbert < \infty $, $\theta$ may cover full range (often called the full model.).

A quantum state for $N$ systems, $\rho^{(N)}$, is described on the $N$-fold tensor product Hilbert space $\Hilbert ^{\otimes N}$.
Suppose that a system composed of $N$+$M$ subsystems is given and that a measurement is performed only for selected $N$ subsystems with the other $M$ subsystems
 left.
Then, the measurement is described by $\{ E_x \otimes I \} $, where $\{ E_x  \} $ is a POVM on $\Hilbert ^{\otimes N}$ and 
$I$ is the identity operator on $\Hilbert^{\otimes M} $.

Our aim is to estimate the true state $\sigma_{\theta }:= \rho_{\theta }^{\otimes M}$ of the remaining $M$ subsystems
 by using a measurement $\{ E_x \}$ on the selected $N$ subsystems
 $\rho_{\theta}^{\otimes N}$. 
We fix an arbitrarily chosen measurement.
Note that the above measurement is not necessarily in the form of a tensor product $E_x^{\otimes N}$,
 which represents a repetition of the same measurement $E_x$ for each system.
Thus, all possible measurements on $N$ subsystems, which may use entanglement,  are considered.
We call $\hat{\sigma}(x)$ an estimate of the true state as {\it a predictive density operator}. 
The problem of the quantum prediction is to seek for the optimal predictive density operator.

The performance of a predictive density operator $\hat{\sigma }(x) $ is evaluated by the quantum $\alpha $-divergence
$D^{(\alpha )}( \sigma_{\theta} || \hat{\sigma} (x) ) $, a quantum analogue of the $\alpha$-divergence (\ref{eq:alphadiv}) in classical statistics.
The quantum $\alpha$-divergence from $\rho $ to $\sigma$ is defined by
\begin{equation}
D^{(\alpha)} (\rho || \sigma) := \frac{4}{1- \alpha^2} \left( 1 - \Tr \sigma^{\frac{1+\alpha}{2}} \rho^{\frac{1-\alpha}{2}} \right), \quad \mbox{if $ \alpha \neq \pm 1$}.
 \label{eq:rel}
\end{equation}
When $\alpha = \pm 1 $, it is defined by
\[
D^{(\alpha=-1)} (\rho || \sigma) := \Tr \rho (\log \rho - \log \sigma) =: D^{(\alpha=1)} (\sigma || \rho).
\]
It satisfies the positivity condition $D^{(\alpha)}(\rho || \sigma ) \geq 0$ and $D^{(\alpha)}(\rho || \sigma) =0 \Leftrightarrow \rho = \sigma$.
Other properties and useful inequalities, see, e.g., \cite{Amari2000}.
Thus, it can be used as a measure for the goodness of a predictive density operator.
Note that the quantum $\alpha$-divergence is reduced to the classical $\alpha$-divergence when the two density operators are commutative.
\\

\noindent
{\bf Remark. 1.}\\
The quantum $\alpha$-divergence can be given a suitable meaning as a measure  only when $|\alpha | \leq 3$. (See, Hasegawa~\cite{Hasegawa1993}.)
However, our statement formally holds for any $\alpha$.
We also assume additional conditions on density operators such that the $\alpha$-divergence is finite.\\
{\bf Remark. 2.}\\
The quantum $\alpha$-divergence can be rewritten in the relative $g$-entropies $H_{g}(\rho || \sigma) := \Tr \rho^{\half} g(L_{\sigma}/R_{\rho})(\rho^{\half} )$,  
where $g$ in an operator convex function and $g(1) = 0$ and $L_{\sigma}(X) = \sigma X, R_{\rho}(X) = X \rho$ are superoperators.
The relative $g$-entropy was introduced by Petz~\cite{Petz1986}.\\





If we assume a prior probability density $\pi(\theta ) $ on the parameter space $\Theta $, the mixture state for the whole $N$ systems is given by
\begin{equation}
 \rho^{(N)} := \int \dtheta \ \pi(\theta) \ \rho_{\theta}^{\otimes N}  \label{eq:exca}.
\end{equation}
A state of the form (\ref{eq:exca}) is called an \textit{exchangeable state}~\cite{Schack2001}, and arises, e.g., if each subsystem is  
prepared in the same unknown way, as in quantum state tomography.
In a quantum exchangeable model~(\ref{eq:exca}), as Schack \textit{et al.}~\cite{Schack2001} showed, 
a posterior distribution $\pi (\theta | x)$ naturally arises.
We assume that the whole system is in the exchangeable state and in our setting  $\pi(\theta |x)$ is given by 
\[
 \pi(\theta | x) := \frac{p(x | \theta)\pi(\theta ) }{\int \dtheta  \ p(x | \theta)\pi(\theta ) },
\]
where $p(x| \theta ) = \Tr \rho_{\theta }^{\otimes N}E_x $.

Finally, let us define generalized Bayesian predictive density operators.
 First of all, we consider an $\alpha$-mixture of $\sigma_{\theta}$ with respect to a posterior density $\pi(\theta | x) $. 
\[
 \sigma^{(\alpha )}_{\pi } (x) := \left\{
 \begin{matrix}
  \{ \int \sigma_{\theta} ^{\frac{1-\alpha}{2}} \pi(\theta | x ) \rmd \theta  \}^{\frac{2}{1-\alpha}}, 
 & \alpha \neq  1, \\
  \exp\left\{  \int \log(\sigma_{\theta}) \pi(\theta | x) \rmd \theta  \right\},   & \alpha = 1.
\end{matrix}
 \right.
\]
Clearly the above mixture is a positive operator and $\Tr \sigma_{\pi}^{(\alpha)}(x) > 0 $.
Thus, we define the generalized Bayesian predictive density operator in the following normalized form.
\[
\tilde{\sigma}_{\pi}^{(\alpha )} (x) := \frac{1}{C_{\alpha}(x)}\sigma_{\pi}^{(\alpha)}(x), \quad C_{\alpha}(x) := \Tr \sigma_{\pi}^{(\alpha)}(x).
\]
In the following section, we show that the generalized Bayesian predictive density operator is the best predictive density operator
 in the sense that it minimizes the averaged quantum $\alpha$-divergence from the true density operator.


\section{Main theorem}

In classical statistics,  Corcuera~and~Giummol\`{e}~\cite{Cor1999} showed that
the generalized Bayesian predictive density $p^{(\alpha)}_{\pi}(y|x)$ is the best predictive density under the $\alpha$-divergence
 when a proper prior $\pi (\theta) $ is given.
We derive the corresponding result for quantum predictive density operators.\\
\\

\noindent
\textit{Theorem.}\\
Let $\alpha \in \real$ be fixed.
Suppose that we perform a measurement for selected $N$ subsystems $\rho_{\theta }^{\otimes N}$of 
 a system $\rho_{\theta }^{\otimes (N+M) }$ composed of $N+M$ subsystems 
in order to estimate the remaining $M$ subsystems $\sigma_{\theta } = \rho_{\theta }^{\otimes M}$. 
The true parameter value $\theta $ is unknown and a prior probability density $\pi (\theta)$ is assumed.
Let $\hat{\sigma}(x) $ be any predictive density operator,
where $x$ is an outcome of a measurement $\{ E_x \}$ for the $N$ subsystems.
Performance of a predictive density operator $\hat{\sigma }(x)$ is measured with the averaged quantum $\alpha $-divergence
\[
 \Exp_{\theta }\Exp_x [D^{(\alpha)}(\sigma_{\theta } || \hat{\sigma }) ]\!\! = \int\! \dtheta \ \pi(\theta )\!\! \int\! \dx \ p(x| \theta)
 D^{(\alpha)}(\sigma_{\theta } || \hat{\sigma }(x) ) 
\] 
 from the true state $\sigma_{\theta }$.
 Then, the generalized Bayesian predictive density operator $\tilde{\sigma}^{(\alpha )}_{\pi}(x)$
is the best predictive density operator.\\
\\
\proof 
When $\alpha \neq \pm 1$, let us consider taking an average of the difference
 between $ D^{(\alpha)}(\sigma_{\theta} || \hat{\sigma} ) - D^{(\alpha)}(\sigma_{\theta} || \tilde{\sigma}^{(\alpha)}_{\pi} ) $.
From now on, we omit $\alpha $ in the $\tilde{\sigma}_{\pi}^{(\alpha)}$.
\begin{eqnarray*}
&&E^{\pi }E^{M_x}[    D^{(\alpha)}(\sigma_{\theta} || \hat{\sigma} ) - D^{(\alpha)}(\sigma_{\theta} || \tilde{\sigma}_{\pi} ) 
 ] \\
&=& \!\!
 \int \! \rmd \theta \pi(\theta) \! 
\int \! \rmd x p(x|\theta) \!
 \left\{ \!
    \frac{4}{1 -\alpha^2} 
 \Tr \sigma_{\theta}^{\frac{1 - \alpha}{2}}  ( \tilde{\sigma}_{\pi}^{\frac{1+\alpha}{2}}  - \hat{\sigma}^{\frac{1+\alpha}{2}} )  
 \! \right\}
 \\   
&=&\!\!\!
\int \!\! \rmd x p_x  \!\! \int \!\! \rmd \theta \frac{\pi(\theta) p(x|\theta)}{p_x}
 \! \left\{ \!
    \frac{4}{1 -\alpha^2} 
 \Tr \sigma_{\theta}^{\frac{1 - \alpha}{2}} \!  ( \tilde{\sigma}_{\pi}^{\frac{1+\alpha}{2}} \! - \! \hat{\sigma}^{\frac{1+\alpha}{2}} )  
 \! \right\}
 \\    
 &=&
 \!\!\int \! \rmd x  p_{x} \!
\int \!\! \rmd \theta \  \pi(\theta | x)
 \! \left\{ \!
    \frac{4}{1 -\alpha^2} 
 \Tr \sigma_{\theta}^{\frac{1 - \alpha}{2}}  ( \tilde{\sigma}_{\pi}^{\frac{1+\alpha}{2}}  - \hat{\sigma}^{\frac{1+\alpha}{2}} )  
 \! \right\}
 \\   
 &=& 
 \!\! \int \! \rmd x p_{x} \frac{4}{1 -\alpha^2} 
 \Tr
\left\{ \! \left( \!\!  \int \!\! \rmd \theta \pi(\theta | x) \sigma_{\theta}^{\frac{1 - \alpha}{2}} \! \right)
  \! ( \tilde{\sigma}_{\pi}^{\frac{1+\alpha}{2}}  \! - \! \hat{\sigma}^{\frac{1+\alpha}{2}} ) \!
 \right\}  
\\    
  &=&
  \int \rmd x p_{x} \frac{4}{1 -\alpha^2} 
\Tr
 \left\{  C_{\alpha}^{\frac{1 - \alpha}{2}} 
   \tilde{\sigma}_{\pi}^{\frac{1 - \alpha}{2}}
  ( \tilde{\sigma}_{\pi}^{\frac{1+\alpha}{2}}  - \hat{\sigma}^{\frac{1+\alpha}{2}} )   
 \right\}
\\     
 &=&
  \int \rmd x p_{x} C_{\alpha}^{\frac{1 - \alpha}{2}} 
\frac{4}{1 -\alpha^2} 
  \left\{   1 -\Tr \tilde{\sigma}_{\pi}^{\frac{1 - \alpha}{2}} \hat{\sigma}^{\frac{1+\alpha}{2}} )    \right\}
\\    
 &=&
 \int \rmd x p_{x}  C_{\alpha}^{\frac{1 - \alpha}{2}} 
D^{(\alpha)}( \tilde{\sigma}_{\pi} || \hat{\sigma} )   \geq 0,
\end{eqnarray*}
where $ p_{x} := \int \rmd \theta' \pi(\theta') p(x | \theta') $ is the marginal density of $x$.
The last inequality holds due to the positivity of the quantum $\alpha$-divergence $D^{(\alpha)}(\sigma || \sigma') \geq 0$ and $p_x \geq 0$.
Since $\hat{\sigma} (x)$ is arbitrarily chosen, it is shown that $\tilde{\sigma}^{(\alpha)}_{\pi}(x) $ is better than any other $\hat{\sigma }(x) $.
We can repeat the same procedure for $\alpha = \pm 1$.\\
\qed

\section{Remarks}

When $\alpha =0$,  the quantum $\alpha$-divergence is closely related to the fidelity $F(\rho, \sigma):= \Tr | \sqrt{\rho} \sqrt{\sigma}|$,
 where $|A|:=\sqrt{A^{*}A} $. Since $|\Tr A | \leq \Tr|A|$, we obtain 
\[
 D^{(0)}(\rho || \sigma) = 4(1-\Tr \sqrt{\rho} \sqrt{\sigma} ) \leq 4(1- F(\rho, \sigma)).
\]
The equality holds when $\rho $ and $\sigma$ are commutative or both $\rho$ and $\sigma$ are pure states.
The fidelity is often used as a measure in the quantum information theory~\cite{Nielsen}.
How our theorem can be extended when we adopt the fidelity as a loss function is left for the future study.


\begin{acknowledgments}
F.T. was supported by the JSPS Research Fellowships for Young Scientists.
\end{acknowledgments}

\newpage 

\end{document}